\documentclass[11pt]{article}
\usepackage{latexsym}
\usepackage{verbatim}
\usepackage{geometry}
\usepackage{graphicx}

\usepackage{soul}

\usepackage{xcolor}
\usepackage[
    colorlinks=true,
    urlcolor=blue,
    linkcolor=blue,
    citecolor=blue,
    filecolor=blue,
]{hyperref}
\usepackage{memhfixc}
\usepackage{epsf}

\usepackage{wrapfig}

\def\realspaces{\catcode`\ =\active}
{\realspaces\global\let =\ }

\usepackage{enumitem}

\setlist[1]{parsep=0pt}
\setlist[1]{itemsep=0pt}

\setlist[2]{parsep=0pt}
\setlist[2]{topsep=0pt}

\makeatletter
\let\@fnsymbol\@arabic
\makeatother

\begin{document}

\title{Learning Outcome Oriented \\ Programmatic Assessment}
\date{1 September 2020}
\author{
Pum Walters
\footnote{dr H.R. (Pum) Walters RI, email: \href{mailto:pum@babelfish.nl}{pum@babelfish.nl}}
~
Michael Nieweg
\footnote{M.R.Nieweg, email: \href{mailto:nieweg.mr@xs4all.nl}{nieweg.mr@xs4all.nl}\href{mailto:mrnieweg@xs4all.nl}{ }}
~
James Watson
\footnote{J.A.Watson, email: \href{mailto:james.watson@bluewin.ch}{james.watson@bluewin.ch}}
\\{\small  Amsterdam University of Applied Sciences}
}

\maketitle

\begin{abstract}
This paper describes considerations behind the organisation of a third semester BSc education. The project aims to facilitate a feedback-oriented environment using assessment for learning and for incremental measure of learner progress [Vleuten {\em et al}, 2012, ``A model for programmatic assessment fit for purpose'']. Learning outcomes encourage higher order cognitive skills, following [Biggs \& Tang, 2011, ``Teaching for quality learning at university: what the student does'']. Embracing [Dochy {\em et al}. 2018, ``Creating Impact Through Future Learning: The High Impact Learning that Lasts (HILL) Model''], several mechanisms encourage focus and motivation.

{\em Key words \& Phrases:} 
Learning Outcomes, Programmatic Assessment, Constructive Alignment, High Impact Learning, Making Learning Outcomes, SOLO Taxonomy, Feedback as Teaching, Assessment for Learning, Action Learning

\end{abstract}




\setcounter{page}{1}
\pagenumbering{arabic}


\section{Introduction}

\label{scrivauto:7}

Traditional methods of teaching seem to be failing (Wolk 2011). In our 2nd-year BSc Cybersecurity education, a number of specific issues were identified. Frontal teaching was becoming less effective: 100 minute attention spans are tough; many students intend to read the slides and books later (but by then time is at a premium); and one-size-fits-all teaching leaves both fast and slow students at a disadvantage. Scheduling multiple classes and multiple projects in a week, and expecting students to move around and even switch teams repeatedly, doesn't stimulate effective work methods. Exams for all classes at the end of each 8 week period, often at awkward times, heightens both study pressure and exam stress. Mechanisms to aid students who need additional guidance, such as on-demand consultancy, were notably ineffective. Overall, motivation and results were dropping.

Many new educational methods are being developed to improve the efficacy of teaching, learning and assessment, e.g. (Sharples et al. 2015) (Ferguson et al. 2019). To mention a few examples: methods which control the learning environment (Crossover learning, Self learning); methods which control the learning activities (Learning Through Argumentation, Learning by doing science with remote labs, Adaptive Teaching, Design thinking, Flip the classroom, Gamification); methods which control assessments (Stealth assessment, Analytics of emotions). (Dunne and Martin 2006) (Fulton 2012) (Mitra and Dangwal 2010)

In this paper we incorporate Learning Outcomes (Adam, 2002), (Kennedy, Hyland \& Ryan, 2002) and Programmatic Assessment (Vleuten et al., 2012),  as a means to increase efficacy. This specific approach shares traits with Crossover Learning, Assessment for Learning, Learning to Learn, Self Learning, Adaptive Teaching, Action Learning and Flip the Classroom. (Sharples et al., 2015), (Ferguson et al., 2019)

In this paper we will refer to this approach as `the project'. The project is now in its second year running and shows increased student activity and productivity, measurable advancement over the semester and enhanced inter and intra-team communication. Perhaps most importantly, the project stimulates both advanced students and students that require more guidance to grow and excel within their own scope. 

Section 2 describes Programmatic Assessment and Section 3 describes Learning Outcomes and their role in the project. Section 4 presents the project in some detail and Section 5 discusses several relevant considerations. Section 6 offers specific guidelines on how to create Learning Outcomes. In Section 7 we briefly address embedding Learning Outcome oriented Programmatic Assessment in an existing context. As far as we are aware, this combination of Programmatic Assessment based on Learning Outcomes, the integration with High Impact Learning, and the process for creating Learning Outcomes using SOLO verbs (Biggs \& Collis, 1982) are new.

\section{Programmatic Assessment}

\label{scrivauto:8}

Programmatic assessment is an integral approach to the design of an assessment program with the intent to optimise its learning function, its decision-making function and its curriculum quality-assurance function (Vleuten, Schuwirth, Driessen, Govaerts \& Heeneman, 2014). We use formative assessments as an important learning activity, and summative assessments to base high-stake decisions on -- i.e. major pass/fail decisions (Vleuten et al., 2012).

\begin{wrapfigure}{r}{240pt}\centering\includegraphics[width=280pt,height=108pt]{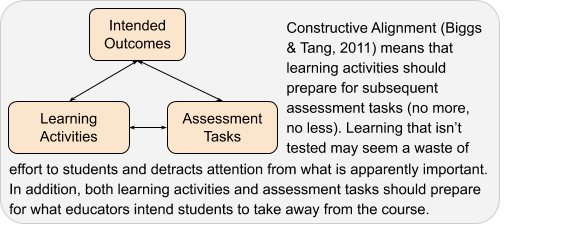}\end{wrapfigure}
The student receives regular and frequent feedback on products and activities, so that both students and teachers have a clear picture of the student's progress. Constructive alignment is very important here: the activities (and products produced in them) and the feedback on them must give the student a substantial basis of preparation for summative tests and of growth with respect to the intended learning results. If constructive alignment is sufficiently high, the feedback itself is an important learning activity.

Following (Vleuten, Schuwirth, Driessen, Govaerts \& Heeneman, 2014), the project offers a program of formative and summative feedback which is an integral part (and the main component) of teaching. Students develop four significant team products and five individual essays. Feedback is offered frequently on their work, their way of working and their team process. This feedback includes weekly professional development coaching, technical progress coaching, incidental written and verbal feedback on intermediate products, and written (formal) feedback on individual essays and team products, and finally, peer feedback. In terms of (Earl, 2014) this is Assessment {\em for} Learning and Assessment {\em as} Learning, and in the end Assessment {\em of} Learning. This feedback includes an indication of to the student's progress with respect to the learning outcomes. The feedback is primarily formative but does have a summative aspect: products and feedback are included in the portfolio that the student builds over the semester. Teachers confer regularly among themselves to objectify feedback and formal feedback on products is provided by two or more teachers. Students that fail to pass on any of the learning outcomes get remedial work aimed specifically at their underdeveloped areas. In principle, feedback is never intended to result in submission of an improved version of the same product. Clearly just processing feedback does not lead to proof at an appropriate professional level. But students can always ask for specific intermediate feedback.

\section{Learning Outcomes}

\label{scrivauto:9}

\begin{wrapfigure}{r}{210pt}\centering\includegraphics[width=250pt,height=99pt]{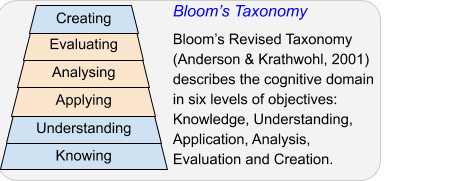}\end{wrapfigure}
Learning outcomes (LO's) could be described as `knowledge students must acquire, understand and be able to reproduce', e.g., (Adam, 2002). A list of definitions of 'learning outcomes' in (Kennedy, Hyland \& Ryan, 2002) shows these key concepts: `Do / Demonstrate', `Knowledge / Understanding', `Attitude'. For a professional, knowledge and understanding are more an essential means than an end; in Bloom's taxonomy, application, analysis and even evaluation characterise professionalism more closely. In Creating Significant Learning Experiences (Fink, 2013), an alternative taxonomy aimed at designing college courses is presented. Simplifying, one might say Bloom's six levels are compressed into Knowledge, Application and Integration, to make space for Human Dimension, Caring and Learning how to Learn. (Biggs \& Tang, 2011) speaks of Intended Learning Outcomes in relation to the constructive alignment mentioned earlier. Finally, we follow (Allan, 1996) who distinguishes subject-based outcomes (complex discipline-based outcomes which are capable of being assessed) from personal transferable outcomes, which one might also call professional skills.

Learning outcomes then, describe a demonstration of application, analysis and evaluation, but importantly, learning outcomes describe behavior and attitude. 

Students in the project aim to be IT professionals with a specialisation in Cybersecurity. Because of this, two of the four learning outcomes address the two core IT processes: requirements-to-design and design-to-implementation. The third addresses Cybersecurity best practice, and the last professional competencies.

Here, we discuss the first LO in detail; the other three appear in Appendix A. 

\begin{quote}\textit{LO 1: A professional, in close communication with the customer, analyses the actual situation and requirements, advises the customer on possible approaches, and iteratively develops an architecture of secure (cloud) information systems and network infrastructure based on accepted principles and patterns. }\end{quote}

This learning outcome is detailed in a rubric to clarify levels of performance on several contained aspects. For instance, in order to demonstrate `adequate' behavior on this learning outcome, the work of a student should be judged at least adequate in its stakeholder and requirements analysis, approach analysis, progressive development and final documentation.

\begin{figure}[htbp]\centering\includegraphics[width=420pt,height=305pt]{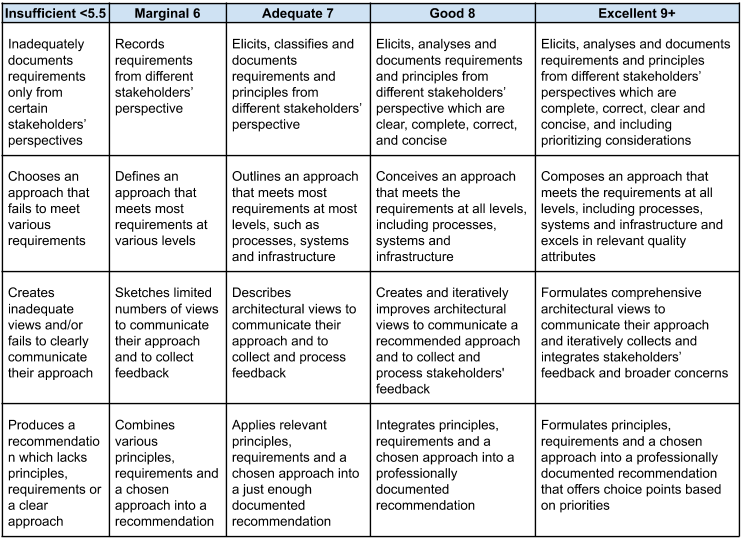}\end{figure}

\emph{Note: even though the LO's are abstract and difficult for students to grasp, the LO rubric is shared with students and a workshop aids initial understanding. After that, the significant formal and informal feedback build a clear mental model (if not actual understanding) of the learning outcomes. This deviates from common practice, where full disclosure and understanding is presumed from the outset. Understanding the end-terms is an essential part of the learning process.}

\section{Project}

\label{scrivauto:10}

\begin{minipage}{\textwidth}

\begin{wrapfigure}{r}{160pt}\centering\includegraphics[width=200pt,height=156pt]{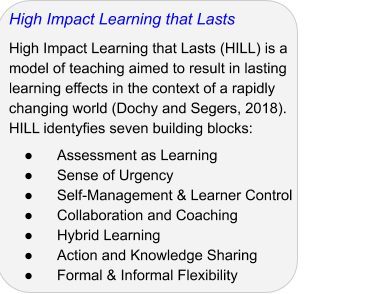}\end{wrapfigure}

The project aims to maximize the individual learning yield for its participants. We are inspired by (Heilesen (ed) \& Andersen (ed), 2015) and (Dochy \& Segers, 2018).

\begin{itemize}
\item The project covers a semester, or to be precise 19 weeks and 24 ECTS (in our sitation the remaining 6 ECTS are spent in generic classes outside the scope of the project).
\item Week one is used for team forming and to choose the subject of the first team product. One purpose is to instill a sense of urgency. Following (Blomh\o j, Enevoldsen, Haldrup \& M\o ller Nielsen, 2015), students are allowed to self-form teams based on self-chosen subjects (in the field of cybersecurity)
\end{itemize}
\end{minipage}

\begin{itemize}
\item Also in week one students get a study manual which contains\begin{itemize} 
\item 	The learning outcomes
\item 	All deadlines
\item 	Rules and guidance governing the project and examination
\item 	Various resources that may be of help \end{itemize}
\item In the sixteen following weeks (disregarding vacation) students work on:\begin{itemize} 
\item 	Formal products (part of the portfolio):\begin{itemize} 
\item 	Five individual essays in which a cybersecurity subject is researched and documented
\item 	Two solution-oriented team products in which a solution is prototyped and documented
\item 	Two customer-oriented team products in which a professional advice is produced
\item 	A project reflection in which significant learning experiences are considered \end{itemize}
\item Informal products and activities (aimed to obtain formative feedback only)\begin{itemize} 
\item 	Two posters aimed to get peer-feedback and ideas
\item 	Several auxiliary products leading to formative feedback (project plan, intermediate versions, plannings, backlogs, meeting notes) \end{itemize}  \end{itemize}
\item On a number of occasions students and teams are required to offer peer feedback to products of other students/teams.
\item Numerous workshops and guest lectures are organised on professional or technical subjects. Participation isn't mandatory, and the workshops are recorded and made available.
\item In the next to last week, students create their reflection report and portfolio (which contains the nine formal products and all feedback obtained).
\item In the final week students have an assessment based on their portfolio, in which the final grade is determined. The assessment is not a technical exam: the teachers and the student are aware of the students professional behavior and technical development. The purpose is:\begin{itemize} 
\item 	To discuss the reflection, and in particular the way in which the student has dealt with feedback `insufficient' for any recent products
\item 	To determine an overall grade based on the performance levels for each learning outcome \end{itemize}
\end{itemize}

\section{Discussion}

\label{scrivauto:11}

A number of important considerations should be mentioned:

\subsection*{Templates}
Our students are 17-to-22-year-olds with a high-school or vocational previous education. Their expected model of teaching is conventional: teachers in front of the class telling students what to do. The project is a culture shock. Having to choose subjects, to invent realistic customer situations, to research approaches, to develop solutions and to produce professional documentation feels awkward to them. Some approach teachers and ask for guidelines, templates, detailed explanations and lectures. However, templates too often become todo-lists; any explanation, prepared table of contents or list of common subjects (e.g. `common stakeholders') invites students to focus on and limit themselves to that list or template, rather than looking at it in a more open-ended way. Even without supplying such materials we have to be careful; many students are happy just to copy the first web search hit as a template.

\subsection*{Curriculum}
The field of Cybersecurity is broad and rapidly changing. By definition, books are likely to be outdated within a couple of years, if not sooner. For the students, many of the underlying principles have been covered in year one; for year two the key outcome is the students' ability to research, analyse and apply new expertise with a professional attitude and behaviour. Material that we cover in workshops and guest lectures, and that we suggest students investigate, is inspired by (CSEC, 2017) and (NIST, 2019).

At the time of writing, control of student's coverage of this material is incidental. Future research includes the development of a skill tree (a graph representation of curricular subjects and their dependencies, such as in (ACM 2017)) and a mechanism to stimulate students to work on obligatory or optional aspects in this skill tree. Such mechanisms might include a system of badges that students acquire doing (formative) tests and assignments.

\subsection*{Optimistically or Not?}
A common model for assessments over the course of a semester is to start out `easy' and progress in complexity and span towards `hard'. That way students perform at an increasingly difficult level. This means that a solution to some specific easy problem might be worth a high grade early in the semester but would be graded lower later on, because the complexity or span is subpar at that point in time. Some teachers also believe that it is important to give students the opportunity to get some high grades early on, to get positive self-confirmation.

When students can almost freely choose subjects, this implies the need to evaluate a product in the context of learning outcomes based on the point in time in the semester. This approach tends to become subjective.

In the project, all feedback, from day one, is given with respect to the LO's as it would be on the final day. In order to avoid negative self-confirmation, a workshop is given to help students interpret these early `bad' grades: at this point in time the qualitative feedback is much more important than an indication of a grade (which is just `failing'; as the learning outcome rubric shows there is only one failing grade: $<$5.5). Experience show students understand this and appreciate receiving an indication of grades at the level of their final test in the semester. In addition this approach serves to heighten the `sense of urgency' that is required and may (with proper guidance) enhance students' intrinsic motivation.

\section{Making Learning Outcomes}

\label{scrivauto:12}

This section describes the stepwise process used to create learning outcomes suitable in the context of programmatic assessment.

These learning outcomes are aimed at integral assessment and are not really suitable for small curricular units such as individual classes or specific subjects. Such smaller units are important but tend to direct students' focus on the short term concern of `passing' subjects, both literally and figuratively. They fragment attention rather than stimulating a holistic view of the learning activities.

Learning outcomes based programmatic assessment stimulate student self management. Guided by teachers, students are themselves able to gauge their level of performance and can make informed choices on how to reach higher levels in a learning outcome rubric.

These learning outcomes are intended for large units of 30 (as in our case) or even 60 ECTS (i.e. an entire year). The learning outcomes are abstract and difficult to grasp for students, so they need time to understand how they should study to grow. 

\subsection*{This Takes Time}
Learning outcomes are not only difficult to understand for students, but also for teachers. Several iterations are needed to arrive at the right set of LO's. Not every teacher involved will have the same idea about the assessment program, the curriculum construction or even their ability to allow students to falter and stumble in a feedback-centric project. A teacher who believes in a stringent curriculum will have difficulty accepting learning outcomes which are primarily aimed at behaviour and attitude. Ideally, the LO's are wrought in a small team in which subject expertise, educational experience and sensitivity to language are combined.

\subsection*{High Level Goals}
Step one is a description of a student at the end of the period (semester or year) in a number of lemmas (sentences). This description should be future-proof and allow for developments in the field. Describe each theme process-oriented and strategically. Focus on {\em why} rather than {\em what}, or indeed {\em how} with attention for both {\em normative} and {\em selective} aspects, in terms of (Walters, 2004). Specific details (the `how') are important but don't have to be taken up in the LO's. `Comprehensively applies professional considerations' goes a long way to substantiate detailed mechanisms in any area.

Make sure that the lemmas cover the entire field of expertise and do not overlap; ideally they should be complementary. They are distinguishable from each other and should be regarded in conjunction; they are inseparable. Ideally, there are three to seven LO's.

It is difficult to express highly abstract LO's. Do not be tempted to include all details students should learn (and will be tested on). Instead, describe professional behavior that can only be exhibited if someone knows, understands and can apply such details. Students work on creating proof of that behavior for half a year or more. They have to be challenged to dive into the details. Long, elaborate and detailed learning outcomes render this superfluous, since students have already been handed the process on a plate.

\subsection*{Refinement}
Often, aspects in the learning outcomes need further refinement, for instance when the LO contains or implies a short list of distinct competences or a container subject. Characterise each of these refinements precisely as that what the course aims at (on this aspect). A capable and industrious student would reach the {\em relational level} in the SOLO taxonomy, as discussed below.

\subsection*{Scales}
For each refinement we will now describe performance at different levels placed in columns under the main learning outcome. There is no specific number of columns, but:

\begin{itemize}
\item Too few may deprive students of the challenge to achieve better. If all one can do is pass or fail, excellent students will loose interest.
\item Too many would blend into an opaque gradient only expressed as a numeric grade.
\end{itemize}

In practice four or five columns make sense. Four columns, for example, might describe `Insufficient', `Marginal', `Good' and `Excellent'.

\begin{wrapfigure}{r}{290pt}\centering\includegraphics[width=350pt,height=230pt]{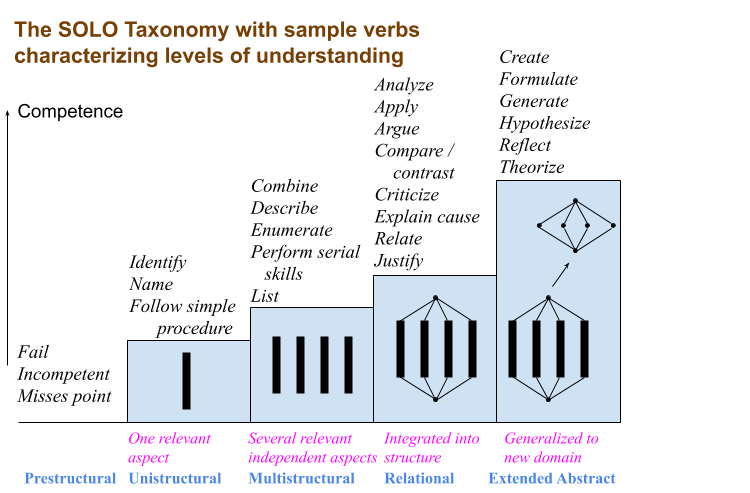}\end{wrapfigure}
The SOLO Taxonomy (Biggs \& Collis, 1982) is developed to categorize levels of understanding in Prestructural, Unistructural, Multistructural, Relational and Extended Abstract. The SOLO taxonomy can be used effectively in formulating learning outcomes, and specific verbs can be associated with different cognitive levels of performance. (Biggs, 2019) shows such a framework, and others can be found searching for `SOLO verbs'.

Following (Biggs \& Collis, 1982) we chose to refine in five columns and present them in the order of increasing levels of understanding. The columns then represent `insufficient', `marginal', `adequate', `good' and `excellent'. Some prefer to order columns from left to right from `Excellent', to `Failing', starting with the `ideal' goal. We are unaware of scientific evidence which order is better; in this paper we adopt the order in the SOLO taxonomy, from `Failing' to `Excellent'.

The descriptions explained under `Refinement' are placed in the column `good' (it is, after all, what we aim for). Now, for each refinement under a learning outcome, we must describe how a student at a level of understanding other than `relational' might perform. The `SOLO verbs' may help us to characterize performance at different cognitive levels. It isn't good enough to simply replace the word `good' by `adequate' or `marginal' because that doesn't help the student to understand, and it leads to a rubric which is subjective.  

It is essential to improve the rubric iteratively, to fine-tune it as a whole and to consider it both horizontally and vertically. Each row must suggest a path of growth for a student. If their performance is marginal, how could they improve it to become good? Each column must `ring true' for an imaginary student that performs at that level of understanding. Many schools and universities offer {\em excellence programs} to motivate and inspire strong students; in our program the column `Excellent' describes the work and attitude of such students.

\subsection*{Last step, confront with Professional Field}
So far, these learning outcomes are cognitive artefacts on the drawing board; one has to be suspicious. It is imperative to validate with employers and workers in the field where students may eventually work. Do the learning outcomes describe their performance? Did we miss or overemphasise things?

\section{Embedding}

\label{scrivauto:13}

Implementing this approach in a standing organization may lead to various concerns.

\subsection*{Students}
\begin{itemize}
\item The Learning Outcomes are very abstract and difficult to understand for second-year BSc students. It is important to advance understanding directly, for instance by offering workshops on LO's, but also indirectly by ensuring that all offered feedback is phrased in terms of the LO's. Students need to understand how to improve their work and thereby get better results with respect to the LO's.
\item Traditional methods of teaching (and often the methods students are used to) involve students sitting back and relaxing while teachers expound on their trade, then cramming in required reading in the last few weeks before an exam. Students may be dissatisfied in a project where they have to work from day one on a vague and very broad curriculum towards learning outcomes they don't understand. 
\item Traditional methods, especially in vocational education, teach by offering templates and to-do lists. Confronted with abstract LO's and a goal of five essays, students typically ask what such an essay must include or if an example of an essay can be given. Sadly, such `help' tends to limit students' own inquisitive research.  
\item Traditional methods could be called \emph{exam-oriented}. In our university, every second-year exam must be offered twice a year, and may be repeated as often as the student is willing to pay tuition. Sadly, this stimulates a mentality where negative feedback can be interpreted as an invitation to `just try again'. In our project there are no retakes on individual products; the only way to recover from a failing grade is to make the next product better, and grades aren't averaged: the final grade reflects students' absolute and relative growth.
\item Students who do not understand or reflect on feedback may simply try to work harder or just try to work in some other way and may feel increasingly dissatisfied with the lack of progress their hard work leads to. Coaching such students is an important success factor of the entire project!
\end{itemize}

\subsection*{Teachers}
\begin{itemize}
\item Offering the right level of feedback takes getting used to. It is in the nature of teachers to want to take the student by the hand and guide them to a higher level, but that may be counterproductive. On the other hand there are few things so sad as a student being stuck for days on end. Talking with students, and with other teachers about all students is needed to identify any sore spots. Weekly team meetings can be instrumental in achieving overview.
\item Many BSc educations distinguish technical teaching from professional skills teaching (as does ours). It is important to narrow this gap: technical feedback should always fit the learning process of the student. Advanced feedback/information may be given, but the student should be able to incorporate this into their own products. 
\end{itemize}

\subsection*{Management}
\begin{itemize}
\item Students, teachers and parents will complain, for all the reasons mentioned above. Management mainly aiming for happy students and parents may get weak at the knees and push for traditional methods or indeed for lowering the bar. It is important that the teacher community consult daily to react --- and `proact' --- in order to keep all students on track, preventing or at least predicting dissatisfied students (and parents).  Management should endorse the renewals described here and should be kept abreast of all results and issues, especially since formative feedback very often doesn't appear in formal grading systems and is therefore invisible outside of the project. Regular management briefings should be incorporated.
\end{itemize}

\section*{Bibliography}

\begin{description}\setlength{\itemsep}{-7pt}\item ACM, IEEE-CS, AIS SIGSEC, \& IFIP WG 11.8. (2017). \textit{Cybersecurity Curricula 2017}. Retrieved from \href{https://dl.acm.org/citation.cfm?id=3184594}{https://dl.acm.org/citation.cfm?id=3184594}\\
\item Adam, S. (2002). Using Learning Outcomes. EUA Bologna Handbook: Making Bologna Work, Volume 1(B2.3-1). Eric Froment (ed), dr Josef Raabe Verlags gmbh, 2006.\\
\item Allan, J. (1996). Learning outcomes in higher education. Studies in Higher Education, 21(1), 93--108. \href{https://doi.org/10.1080/03075079612331381487}{https://doi.org/10.1080/03075079612331381487} \\
\item Anderson, L. W. \& Krathwohl, D. R. (2001). A taxonomy for learning, teaching, and assessing: a revision of Bloom's taxonomy of educational objectives. New York: Longman.\\
\item Berkel, H. van, Bax, A. \& Joosten-ten Brinke, D. (2017). Toetsen in het hoger onderwijs. Retrieved from \href{https://www.bsl.nl/shop/toetsen-in-het-hoger-onderwijs-9789036816786.html}{https://www.bsl.nl/shop/toetsen-in-het-hoger-onderwijs-9789036816786.html} \\
\item Biggs, J. B. (2019). SOLO Taxonomy. Retrieved from John Biggs website:\\ \href{https://www.johnbiggs.com.au/academic/solo-taxonomy/}{https://www.johnbiggs.com.au/academic/solo-taxonomy/} \\
\item Biggs, J. B. \& Collis, K. F. (1982).\textit{Evaluating the quality of learning: the SOLO taxonomy (structure of the observed learning outcome)}. New York; Toronto: Academic Press.\\
\item Biggs, J. B. \& Tang, C. (2011). \textit{Teaching for quality learning at university: what the student does}. Maidenhead, England; New York: Mcgraw-Hill, Society For Research Into Higher Education \& Open University Press.\\
\item Blomh\o j, M., Enevoldsen, T., Haldrup, M. \& M\o ller Nielsen, N. (2015). The Bachelor Programmes and the Roskilde Model. In \textit{ROSKILDE MODEL: problem-oriented learning and project work.} Springer International.\\
\item CSEC. (2017). Curriculum Guidelines for Post-Secondary Degree Programs in Cybersecurity. In \textit{Sybersecurity Curricular Guidelines, CSEC 2017}. Retrieved from ACM, IEEE, AIS SIGSEC, IFIP WG 11.8 website: \href{http://cybered.acm.org/}{http://cybered.acm.org/}. A Report in the Computing Curricula Series Joint Task Force on Cybersecurity Education.\\
\item Dannefer, E. F. \& Henson, L. C. (2007). The Portfolio Approach to Competency-Based Assessment at the Cleveland Clinic Lerner College of Medicine. \textit{Academic Medicine}, \textit{82}(5), 493--502. \href{https://doi.org/10.1097/acm.0b013e31803ead30}{https://doi.org/10.1097/acm.0b013e31803ead30} \\
\item Dochy, F. \& Segers, M. (2018). \textit{Creating Impact Through Future Learning The High Impact Learning that Lasts (HILL) Model} (1st ed.). Routledge.\\
\item Dunne, D. \& Martin, R. (2006). Design Thinking and How It Will Change Management Education: An Interview and Discussion. \textit{Academy of Management Learning \& Education}, \textit{5}(4), 512--523. \href{https://doi.org/10.5465/amle.2006.23473212}{https://doi.org/10.5465/amle.2006.23473212}\\
\item Earl, L. M. (2014). \textit{Assessment as learning: using classroom assessment to maximize student learning}. Hawker Brownlow Education.\\
\item Ferguson, R., Coughlan, T., Egelandsdal, K., Gaved, M., Herodotou, C., Hillaire, G., {\ldots} Whitelock, D. (2019). \textit{Exploring new forms of teaching, learning and assessment, to guide educators and policy makers}. Retrieved from\\\href{https://iet.open.ac.uk/file/innovating-pedagogy-2019.pdf}{https://iet.open.ac.uk/file/innovating-pedagogy-2019.pdf}\\
\item Fink, L. D. (2013). \textit{Creating significant learning experiences: an integrated approach to designing college courses}. San Francisco: Jossey-Bass.\\
\item Fulton, K. (2012). Upside down and inside out: Flip Your Classroom to Improve Student Learning. \textit{Learning \& Leading with Technology}, \textit{39}(8), 12--17. Retrieved from \href{https://eric.ed.gov/?id=EJ982840}{https://eric.ed.gov/?id=EJ982840} \\
\item Heilesen (ed), S. B. \& Andersen (ed), A. S. (2015). \textit{The Roskilde model: problem-oriented learning and project work}. Cham: Springer.\\
\item Kennedy, D., Hyland,  \'{A}. \& Ryan, N. (2002). Writing and using learning outcomes: a practical guide. \textit{EUA Bologna Handbook: Making Bologna Work}, \textit{Volume 1}(C 3.4-1). \href{https://doi.org/9780955222962}{https://doi.org/9780955222962} \\
\item Mitra, S. \& Dangwal, R. (2010). Limits to self-organising systems of learning-the Kalikuppam experiment. \textit{British Journal of Educational Technology}, \textit{41}(5), 672--688. \href{https://doi.org/10.1111/j.1467-8535.2010.01077.x}{https://doi.org/10.1111/j.1467-8535.2010.01077.x} \\
\item NIST. (2019, October 27). Cybersecurity. Retrieved from NIST/Cybersecurity website: \href{https://www.nist.gov/topics/cybersecurity}{https://www.nist.gov/topics/cybersecurity} \\
\item Ryan, M. (2013). The pedagogical balancing act: teaching reflection in higher education. \textit{Teaching in Higher Education}, \textit{18}(2), 144--155.\\\href{https://doi.org/10.1080/13562517.2012.694104}{https://doi.org/10.1080/13562517.2012.694104} \\
\item Ryan, M. (2015). Improving reflective writing in higher education: a social semiotic perspective. \textit{Teaching in Higher Education}, \textit{16}(1), 99--111. Retrieved from \href{https://%5C%5Cwww.academia.edu/3155436/}{https://\\www.academia.edu/3155436/} Improving\_reflective\_writing\_in\_higher\_education\_a\\\_social\_semiotic\_perspective\\
\item Scherer, M. (2016). \textit{ON FORMATIVE ASSESSMENT: readings from educational leadership (el essentials).} Place Of Publication Not Identified], Assoc For Supervision.\\
\item Sharples, M., Adams, A., Alozie, N., Ferguson, R., Fitzgerald, E., Gaved, M., {\ldots} Means, B. (2015). \textit{Innovating Pedagogy 2015 Exploring new forms of teaching, learning and assessment, to guide educators and policy makers}. Retrieved from \href{http://proxima.iet.open.ac.uk/public/innovating%5C_pedagogy%5C_2015.pdf}{http://proxima.iet.open.ac.uk/public/innovating\_pedagogy\_2015.pdf} \\
\item Vleuten, C. P. M. van der, Schuwirth, L. W. T., Driessen, E. W., Dijkstra, J., Tigelaar, D., Baartman, L. K. J. \& van Tartwijk, J. (2012). A model for programmatic assessment fit for purpose. \textit{Medical Teacher}, \textit{34}(3), 205--214. \\
\href{https://doi.org/10.3109/0142159x.2012.652239}{https://doi.org/10.3109/0142159x.2012.652239} \\
\item Vleuten, C. P. M. van der, Schuwirth, L. W. T., Driessen, E. W., Govaerts, M. J. B. \& Heeneman, S. (2014). Twelve Tips for programmatic assessment. \textit{Medical Teacher}, \textit{37}(7), 641--646. \href{https://doi.org/10.3109/0142159x.2014.973388}{https://doi.org/10.3109/0142159x.2014.973388} \\
\item Walters, H. R. (2004). Structuring professional cooperation. Information and Software Technology, 46(6), 415--421. \href{https://doi.org/10.1016/j.infsof.2003.08.004}{https://doi.org/10.1016/j.infsof.2003.08.004}  \\
\item Wolk, R. A. (2011). \textit{Wasting minds: why our education system is failing and what we can do about it}. Alexandria, Virginia: Ascd.\\\end{description}

\section*{Appendices}
\label{scrivauto:15}
\appendix

\section{Learning Outcomes 3rd semester Cybersecurity}

\label{scrivauto:16}

This section describes the intended learning outcomes of the third semester in the Cybersecurity program in the HBO-ICT Bachelor education at the University of Applied Sciences in Amsterdam.

From the perspective of a (Cybersecurity) professional, the learning outcomes describe the effort of aiding a customer with an IT wish to bring about a solution that fulfils their needs. The process and the solution can be professionally substantiated by looking at quality aspects such as security.

In this semester students acquire knowledge and develop skills at a junior level needed to:

\begin{itemize}
\item 	•	Apply Architectural Methods to improve analysis, communication and maintainability
\item 	•	Apply (various) Quality Attributes to safeguard system and process requirements
\item 	•	Implement solutions for information systems within specific time constraints
\item 	•	Produce well documented solutions which meet the customer's goals
\end{itemize}
The learning outcomes are explained in the following process segments. Each segment describes one aspect of the customers' problem-to-solution paradigm which is related to the HBO-i competence framework. The header of each segment defines the process. The cells elaborate this in greater detail, using key concepts, the academic achievement and the professional knowledge, skills and attitude. The column `Good' represents the intended learning outcomes of the course. It is very important to understand this column in depth, because it is required to offer peers useful feedback and to apply reflection in order to improve the quality of one's own work. To the left and right of this column different levels of the achievements are described.

Students -- and professionals -- can apply the scales at any specific time to evaluate their performance. After obtaining feedback, their personal learning goals can be adjusted. Eventually, these scales are also used for the final assessment, at which time each student must at least perform at the `Marginal' level in all categories in order to pass.

\begin{figure}[htbp]\centering\includegraphics[width=420pt,height=280pt]{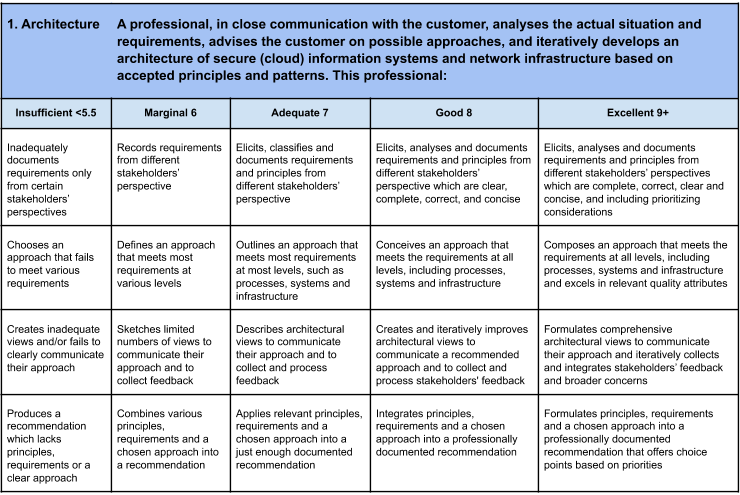}\end{figure}

\begin{figure}[htbp]\centering\includegraphics[width=420pt,height=277pt]{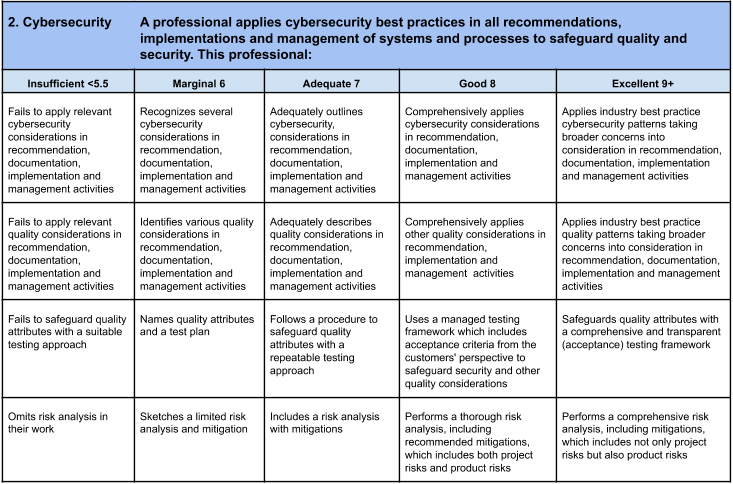}\end{figure}

\begin{figure}[htbp]\centering\includegraphics[width=420pt,height=244pt]{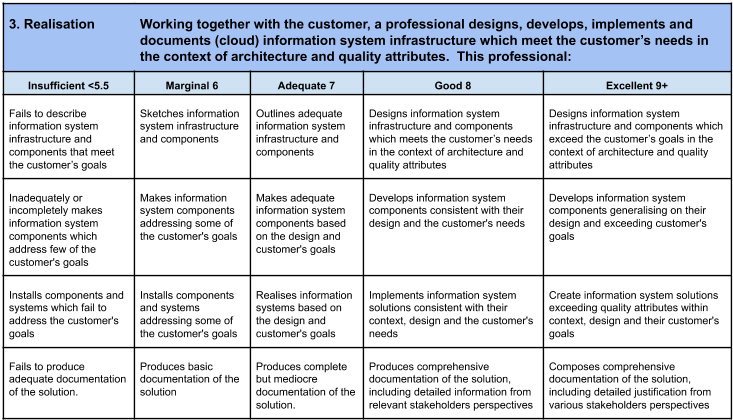}\end{figure}

\begin{figure}[htbp]\centering\includegraphics[width=420pt,height=362pt]{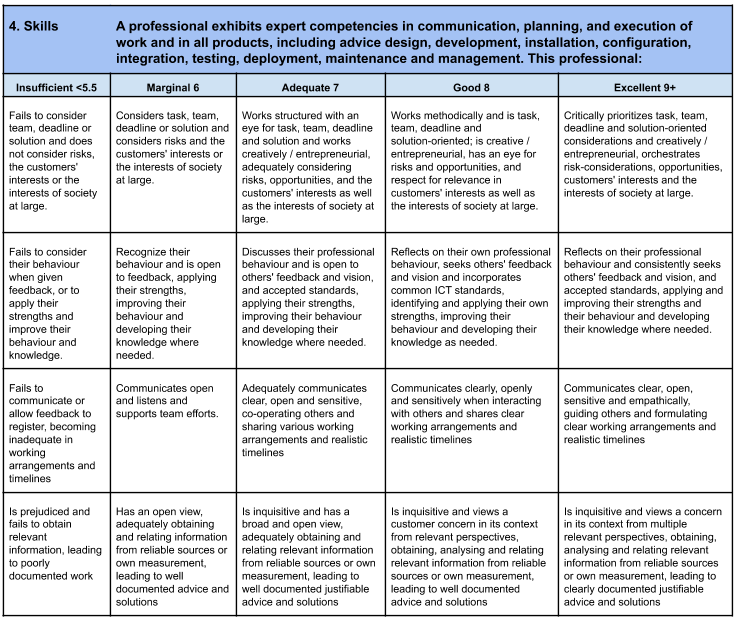}\end{figure}
\end{document}